# Multisensor fusion-based digital twin in additive manufacturing for in-situ quality monitoring and defect correction

Lequn Chen[1,2], Xiling Yao[1*], Kui Liu[1], Chaolin Tan[1], Seung Ki Moon[2*]

[1] Singapore Institute of Manufacturing Technology, A*STAR, Singapore

[2] School of Mechanical and Aerospace Engineering, Nanyang Technological University, Singapore

chen1470@e.ntu.edu.sg (L. Chen); yaox@outlook.com (X. Yao); skmoon@ntu.edu.sg (S.K. Moon)

**Abstract**

Early detection and correction of defects are critical in additive manufacturing (AM) to avoid build failures. In this paper, we present a multisensor fusion-based digital twin for in-situ quality monitoring and defect correction in a robotic laser direct energy deposition process. Multisensor fusion sources consist of an acoustic sensor, an infrared thermal camera, a coaxial vision camera, and a laser line scanner. The key novelty and contribution of this work are to develop a spatiotemporal data fusion method that synchronizes and registers the multisensor features within the part's 3D volume. The fused dataset can be used to predict location-specific quality using machine learning. On-the-fly identification of regions requiring material addition or removal is feasible. Robot toolpath and auto-tuned process parameters are generated for defecting correction. In contrast to traditional single-sensor-based monitoring, multisensor fusion allows for a more in-depth understanding of underlying process physics, such as pore formation and laser-material interactions. The proposed methods pave the way for self-adaptation AM with higher efficiency, less waste, and cleaner production.

(Abstract max: 1200 characters. Currently: 1194; maximum 5 Keywords)

# 1 INTRODUCTION

Additive manufacturing (AM), also known as 3D printing, has shown remarkable potential in manufacturing geometrically complex products with improved mechanical performance, reduced weight, unique functionality, and shortened product design and development lifecycles (Moon et al., 2014; Yao et al., 2015; Tan et al., 2021). In particular, AM-enabled on-demand production of critical components improves capabilities in mitigating risks associated with supply chain interruptions induced by global crises, such as the current long-term COVID-19 pandemic (Choong et al., 2020). However, maintaining high quality consistency, dimensional accuracy, and process repeatability remain a significant challenge, especially for large-format metal AM techniques such as laser direct energy deposition (LDED). To prevent build failures, defects such as porosity, cracking, and geometric distortions must be detected and corrected early during the process.

Recently, significant effort has been made to Artificial Intelligence (AI)-assisted defect detection in AM (AbouelNour and Gupta, 2022). The state-of-the-art in-process sensing technologies (visual, acoustic, thermal, etc.) show promise in predicting specific types of defects (e.g., lack-of-fusion pores, balling, dilutions) or mechanical properties (e.g., tensile strength) for online monitoring of the laser AM process (Drissi-Daoudi et al., 2022). However, because most sensors could not adequately capture the complex melt pool metallurgical mechanism, no single sensing approach could predict defects and part quality holistically. Multisensor monitoring allows for a more in-depth understanding of complicated underlying physical phenomena that were previously unexplored.

In this paper, we propose a framework for location-dependent quality prediction and self-adaptive defect correction in a robotic LDED process. The key novelty and contribution are to develop a novel multisensor fusion-based digital twin that spatiotemporally synchronizes and registers the multimodal features within the part's 3D volumetric domain. Multisensor fusion sources consist of an acoustic sensor, a shortwave infrared (SWIR) thermal camera, and a coaxial vision camera. Real-time robot tool-centre-point (TCP) positions are acquired from a robot controller. After fusing the multisensor features, the multimodal dataset can be used for location-specific quality prediction via machine learning. Multiple quality values (porosity, micro-harness, geometric deviations, etc.) at different locations within the part's volume can be predicted. Based on the 3D quality prediction outcomes, on-the-fly identification of regions requiring material addition or removal is feasible. Decisions can be made on subtractive removal of defective regions or additive restoration of dimensional accuracy. For defect correction, the robot toolpath with auto-tuned process parameters can be automatically generated.

In contrast to traditional single-sensor-based monitoring, multisensor fusion enables a more in-depth understanding of underlying physical events for in-situ defect prediction. The proposed multisensor fusion-based digital twin can improve monitoring reliability by overcoming the limitations of individual sensing sources. The proposed framework paves the way for self-adaptation AM in Cyber-Physical Production Systems (CPPSs) with higher efficiency, less waste, and cleaner production.

# 2 METHOD OVERVIEW

## 2.1 Overview: multisensor in-situ quality monitoring and defect correction framework

Figure 1 shows an overview of the proposed framework for multisensor in-situ quality monitoring and self-adaptive defect correction, which consists of five major steps:
- An in-situ monitoring system with multisensor data sources has been developed. Sensor-captured process data in robotic LDED include coaxial melt pool images, temperature field, audio signals, and 3D point cloud of part surface. Details on multisensor setup are illustrated in Section 2.2.
- Spatiotemporal data fusion is performed to synchronize and register the multisensor features in the part's volumetric domain, which is the key contribution of this work. It is the prerequisite for subsequent quality prediction and defect corrections. Details on the data fusion method is described in Section 3.

- A machine learning (ML)-based approach is proposed to map the spatiotemporally fused datasets to the quality values within the entire volumetric domain. Surface inspection and interior quality inspection are conducted to measure the dimensional deviation, internal defects (porosities, cracking), microstructural features, and mechanical properties (microhardness) of the part.
- Once the ML models are trained, the models will be used to identify regions requiring material addition or removal in online quality prediction. Defect boundaries can be extracted automatically for subsequent self-adaptive defect correction.
- Hybrid process sequences can be auto-tuned to adaptively improve the part quality. Robot toolpath for machining or LDED with adjusted parameters can be automatically generated and executed from the in-house developed software platform, which removes the defects and restores the dimensional accuracy to ensure successful AM production.

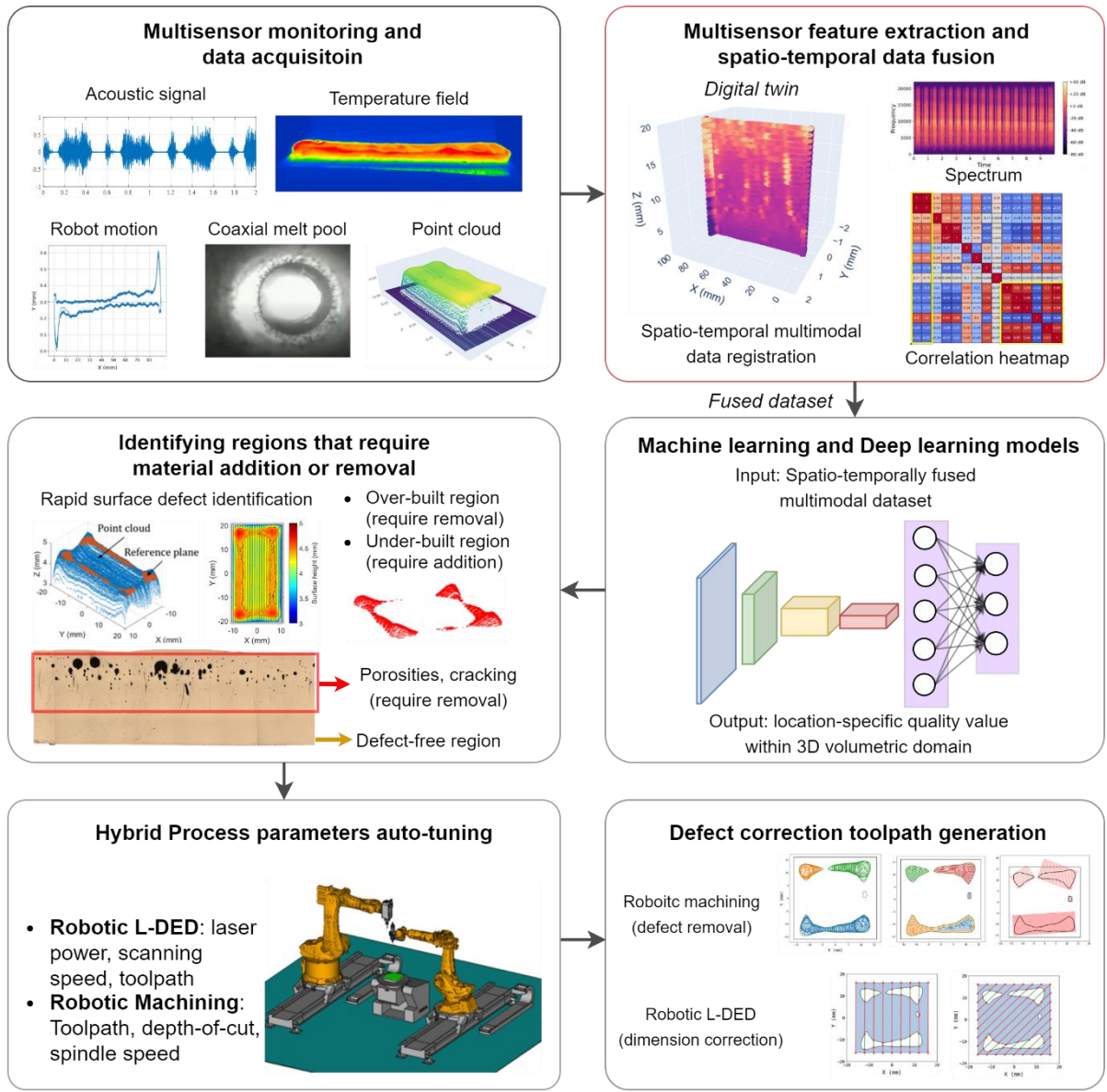

*Figure 1. The proposed framework for multisensor digital twin in robotic LDED AM for in-situ quality monitoring and self-adaptive defect correction.*

## 2.2 System setup

Figure 2(a) depicts the dual-robot hybrid additive-subtractive manufacturing system from the Singapore Institute of Manufacturing Technology (SIMTech). The system consists of a robotic LDED cell and a robotic machining cell. For the LDED process, a coaxial powder-blown nozzle is carried by a 6-axis industrial robot, and the workpiece is held by a 2-axis positioner. A high-energy laser beam with a wavelength of 1070 nm melts powder material as they are deposited onto the substrate. As the nozzle moves in the feed direction, the molten material solidifies rapidly in the molten pool area. For the robotic machining process, the milling spindle or grinding wheel can be mounted on the end-effector of the industrial robot, which is used for the subtractive removal of defective parts. The high degree-of-freedom (DoF) dual-robot hybrid system enables flexible fabrication of extremely large workpieces. As shown in Figure 2(b) and Figure 3, various sensors are integrated into the robotic LDED for in-situ process monitoring:

- An IR thermal camera is installed next to the laser nozzle to monitor the temperature field around the melt pool heat affected zones at a frequency of 120 Hz. Key temperature features such as peak temperature and temperature variances are extracted.
- On the optical head was a coaxially mounted CCD camera with an acquisition frequency of 30 Hz. The melt pool image can be acquired coaxially by the visible spectrum CCD camera through a set of reflecting lenses. An optical NIR band-pass filter was attached to the camera lens to isolate the melt pool from the surroundings. The CCD camera with a NIR optical filter was used to monitor the melt pool morphologies.
- A low-cost microphone sensor is used for monitoring the laser-material interaction sound during the LDED process. The microphone has a frequency response range of 50 – 20000 Hz and was positioned near the powder feeding nozzle, and the sampling rate was set to 44100 Hz. The sound contains environmental noise, which can be reduced by an acoustic denoising approach (Chen et al., 2022).
- Robot motion (TCP position, velocity, acceleration, etc.) is acquired by TCP/IP ethernet communication between the robot controller and the PC. The joint position is measured by the built-in servo encoder in the 6-axis KUKA robot at a frequency of 250 Hz.
- On-machine measurement for in-process surface monitoring was achieved by using the laser line scanning technique (Chen, Yao, Xu, et al., 2020a, 2020b; Xu et al., 2022). A laser displacement sensor was installed on the robot end effector adjacent to the laser head to generate a 3D point cloud of the part surface, which can be used to inspect part surface conditions.

All the sensing modalities are connected to a PC running the Ubuntu Linux Operating system with an in-house developed software platform using Robot Operating System (ROS). The details on the software architecture and proposed data acquisition methods will be shown in section 2.3.

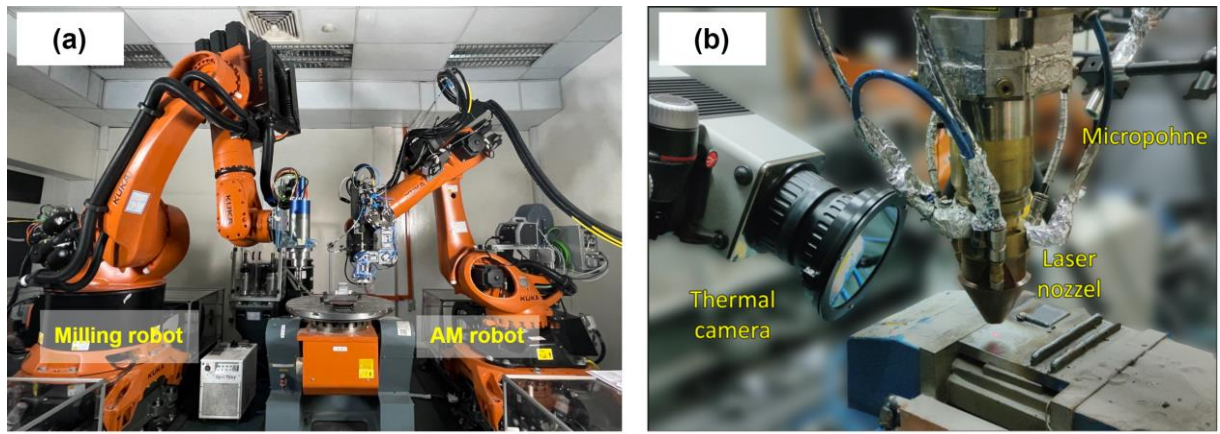

*Figure 2. (a) Dual-robot hybrid additive-subtractive manufacturing system from SIMTech; (b) multisensor monitoring setup for robotic LDED.*

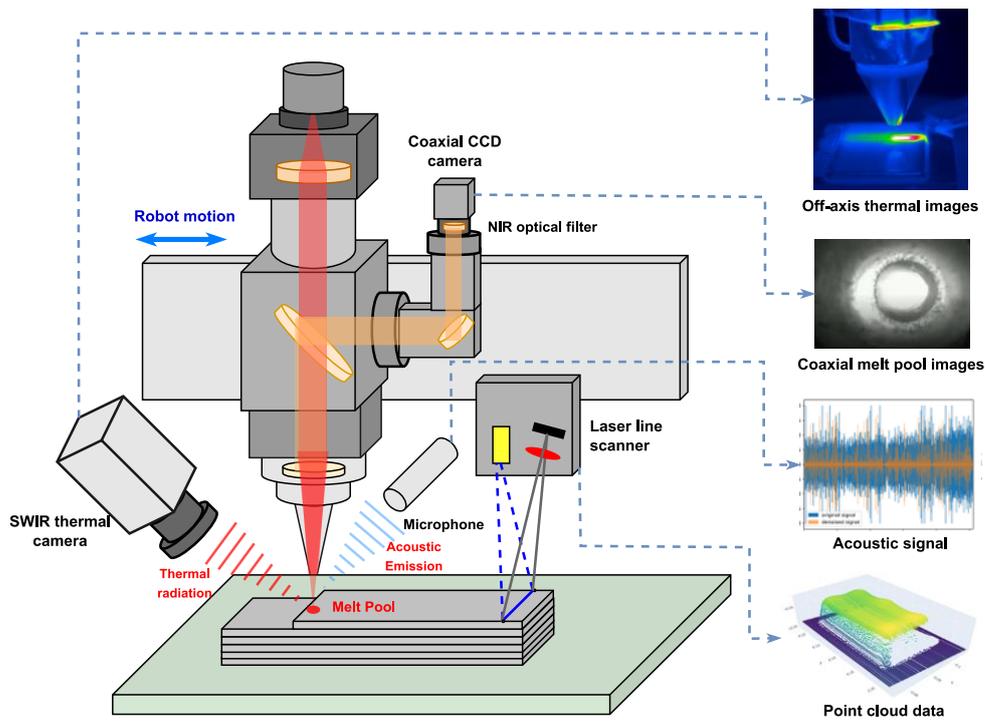

*Figure 3. The schematic diagram of multisensor monitoring setup for robotic LDED system.*

## 2.3 Software Architectures

A multi-nodal software platform is developed in-house to implement the proposed framework. Figure 4 illustrates the software architecture, which was built on top of the ROS melodic (Quigley et al., 2009). As the raw data are retrieved, further data processing and feature extraction procedures, such as acoustic signal denoising, noise filtering, and visualisation are conducted simultaneously in separate nodes. Key sensor data features, such as acoustic spectral descriptors, melt pool morphologies (width, length, contour area, etc.), and temperature can be visualised in real-time.

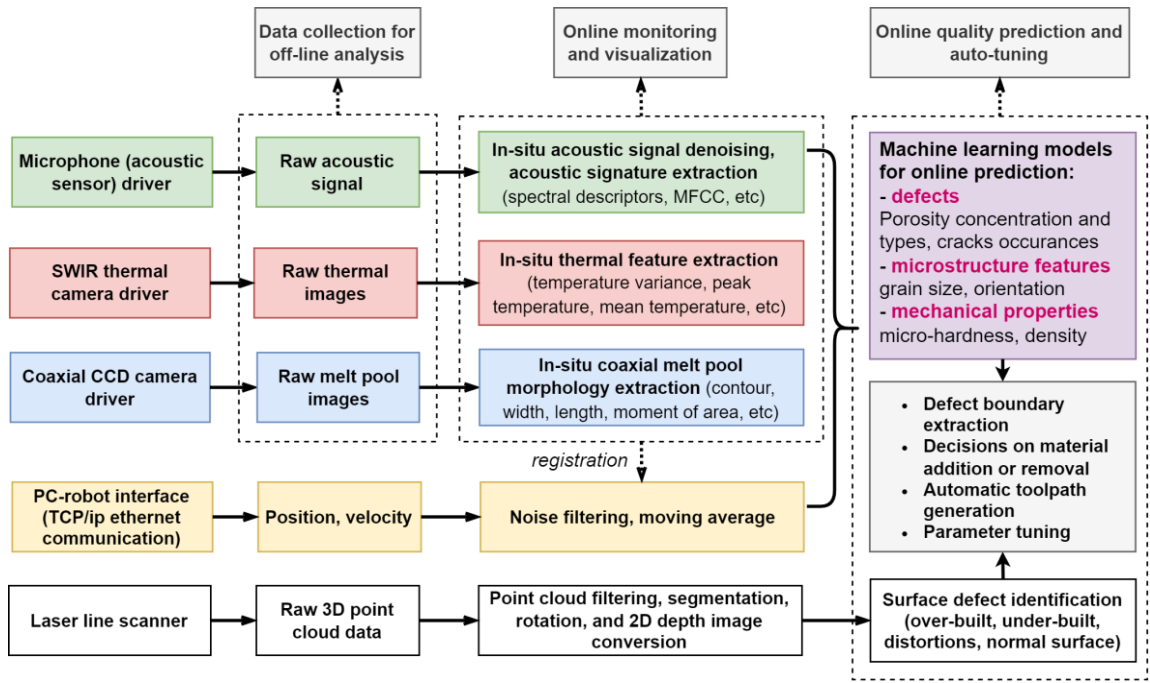

*Figure 4. Software architecture for multimodal monitoring, transfer learning, online quality prediction and self-adaptive quality enhancement.*

The in-situ acquired multisensor feature is utilised for multiple purposes in offline analysis, including: (1) spatiotemporal data fusion for location-specific quality prediction (which will be introduced in Section 3); (2) temperature features collected from SWIR thermal camera with the low acquisition frequency can be predicted by the other two low-cost sensing techniques (e.g., acoustic data and coaxial vision data), which is our ongoing work; (3) transfer learning across different sensing modalities to improve the defect prediction performance of individual sensors (which is our future work). After offline training, the ML models can be deployed into the software pipeline for in-situ quality prediction without process interruption or manual assistance.

The surface topography of the components in the intermediate layers is inspected in-process using laser line scanning, as described in our previous work (Xu et al., 2022). Raw point cloud data contains unwanted substrate surfaces and noise, which are filtered and removed (Chen, Yao, Xu, et al., 2020b). Over-built and under-built surface defects can be recognised and visualised on-the-fly from the 3D point cloud. The online quality prediction outcome is used for identifying regions for defect correction. A robotic toolpath is automatically generated for AM repair or SM removal of identified defects. Hybrid process auto-tuning, automatic toolpath generation and parameter tuning are future work, which will be briefly introduced in Section 4.

## 3 MULTISENSOR DATA FUSION

In this section, we introduce the multisensor feature extraction and spatiotemporal fusion technique for constructing a digital twin, which is the main contribution of this work. The proposed multisensor fusion-based digital twin set the foundation for location-specific quality prediction and defect correction.

### 3.1 Multisensor monitoring and feature extraction

Multiple features are extracted from each sensing modality to provide different perspectives of LDED process, which are illustrated below.

#### 3.1.1 Coaxial melt pool monitoring

The melt pool geometric features can represent the melt pool heat transfer state and process stability. Variations in the melt pool geometry could indicate localised heat accumulation, potential build anomalies and defects (Chen, Yao, Chew, et al., 2020). We adopted a similar method as presented by Knaak et al. (2021) to extract melt pool morphological information using OpenCV (Bradski, 2000) with ROS. As shown in Figure 5, the raw melt pool image (Figure 5(a)) is binarized (Figure 5(b)), and the melt pool contour area is calculated as:

$$m_{00} = \sum_x \sum_y I(x,y) \Delta A \tag{1}$$

which is the 0th-order moment of melt pool pixels, and $I(x,y)$ represents pixel intensities. Based on the contour area, a convex hull can be extracted, as shown in Figure 5(f), which represents the smallest convex point set that contains the melt pool contour. The melt pool contour is fitted by an ellipse in a least square sense, where ellipse width and length denote melt pool width and length, respectively. The elliptical shape can be represented by the following equation:

$$\frac{(x\cos a + y\sin a)^2}{a^2} + \frac{(x\sin a - y\cos a)^2}{b^2} = 1 \tag{2}$$

where $a$ and $b$ represent the ellipse width and length. Melt pool central moments are defined as follows:

$$\mu_{ji} = \sum_x \sum_y I(x,y) \cdot (x-\overline{x})^j \cdot (y-\overline{y})^j \tag{3}$$

where $(\overline{x}, \overline{y})$ is the centre of gravity (COG) of melt pool pixels. The time-domain plots for melt pool moments, width, and length when printing a thin-wall structure are shown in Figures 5(g)-(h). The melt pool central moments are highly sensitive to process anomaly, as seen in Figure 5(g). When depositing in the defective regions, significant variations can be observed. Melt pool ellipse width and length increase over time, as shown in Figure 5(h). The poor heat conduction capability and localised heat accumulation cause the melt pool to become less stable over time as the printing process proceeds. A clear threshold of the melt pool transiting from the stable to the unstable zone can be seen on the melt pool central moment plot.

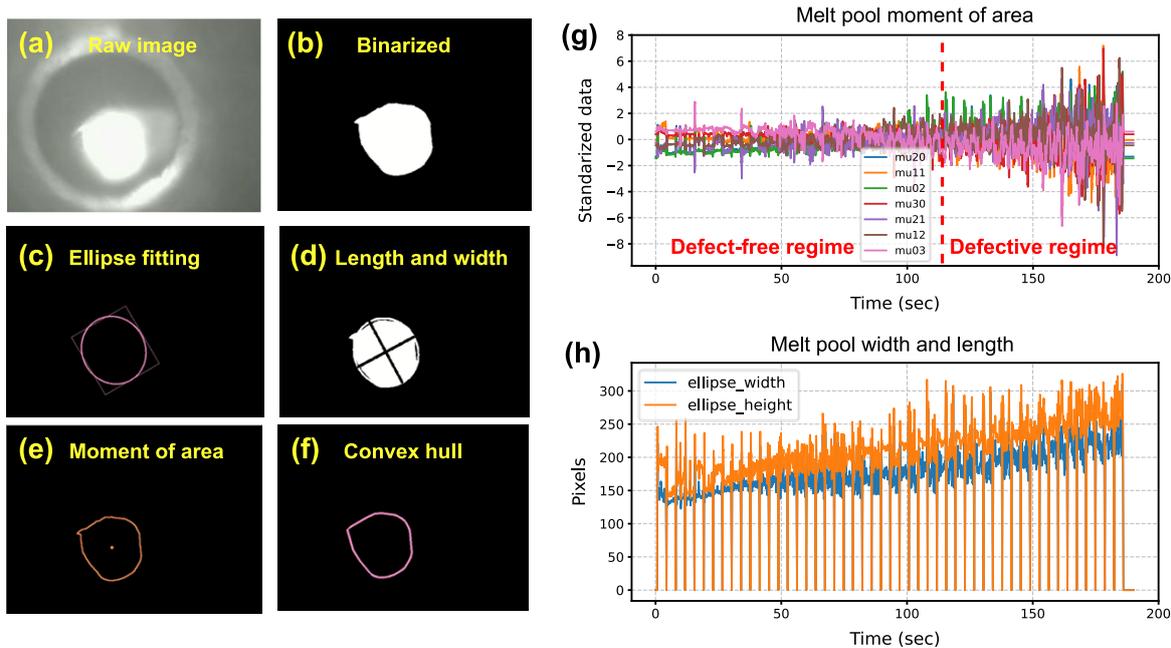

*Figure 5. Coaxial melt pool feature extractions: (a)-(f) coaxial melt pool image processing for morphological feature extraction; (g) time-domain plot of melt pool moment of area when printing a thin-wall structure; (h) time-domain plot of melt pool width and length when printing a thin-wall structure.*

### 3.1.2 In-situ acoustic monitoring

Deploying vision-based monitoring solutions is often time-consuming and expensive. Acoustic-based monitoring methods provide distinct advantages, such as adjustable sensor setup and lower hardware costs. Drissi-Daoudi et al. (2022) reported recent findings that demonstrated the significant potential of employing acoustic signals to distinguish different process regimes in the L-PBF process. The acoustic signal in the DED process, on the other hand, is often noisier, making it difficult to analyse the laser-material interaction sound directly. Recently, we presented an acoustic denoising technique that reduces the environmental noise from the raw acoustic signal (Chen et al., 2022). The denoised signal is used to extract features and perform data fusion analysis. Figure 6 depicts key acoustic signatures collected and plotted from DED experiments with different process parameters.

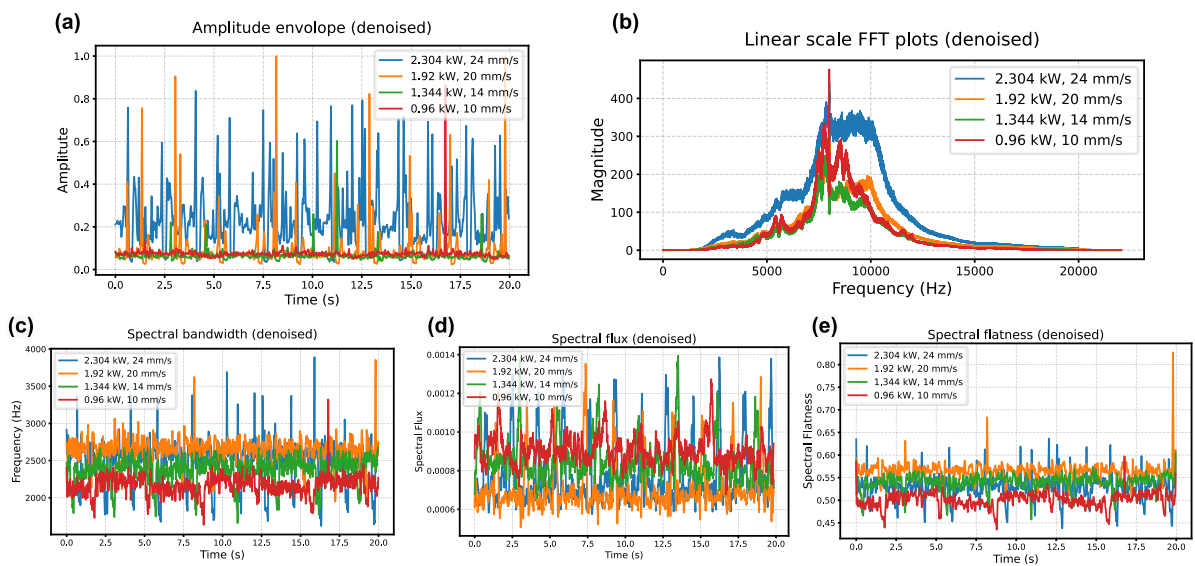

*Figure 6. Visualisation of acoustic features in time-domain and frequency-domain.*

The maximum amplitude value of all samples in a frame is used to calculate the amplitude envelope (AE). The AE feature, as shown in Figure 6(a), can display how acoustic energy fluctuates over time and discriminate sound from different process regimes. The Fast Fourier Transform (FFT) plots also illustrate the distinguishable signal content across the frequency domain for different processes. Three spectral descriptors (Klapuri and Davy, 2006) are plotted in Figure 6(c)-(e). The spectral centroid (SC) is the centre of gravity (COG) of the magnitude spectrum. The spectral bandwidth (SBW) determines the magnitude spectrum variation from the SC. The frequency point where less than 85 per cent of the total energy exists is measured by spectral roll-off (SR). All three spectral descriptors in Figure 6 clearly distinguish the processes, showing the viability of using acoustic signals for in-situ process monitoring. Apart from the spectral descriptors, Mel-frequency cepstrum coefficients (MFCCs) (Muda et al., 2010) are also extracted. The MFCCs takes the inverse Fourier transform of a logarithm of the spectrum of the signal, which can be expressed by the following equation:

$$C_{(x(t))} = F^{-1}[\log(F[x(t)])] \tag{4}$$

where function $C_{(x(t))}$ computes the cepstrum of a signal $x(t)$. $F$ represents the Fourier transform function, and $F^{-1}$ is inverse Fourier transform. The number of MFCCs for each time frame is set to 20, which equally divides the frequency bands into 20 segments. Spatiotemporal variations of MFCCs feature will be shown in section 3.2.

### 3.1.3 Temperature field monitoring

Thermal history and temperature field are also helpful in identifying potential process anomalies and predicting part quality. Before extracting the melt pool temperature feature, emissivity calibration is necessary for IR thermal cameras to ensure correct temperature readings. The emissivity value for the commercial C300 maraging steel is set to 0.3 for the liquid molten pool region and 0.5 for the rest of the head affected zone (HAZ). Figure 7(a) depicts the melt pool and HAZ after adaptively correcting the emissivity. Figure 7(b) shows thermal monitoring when fabricating a thin-wall structure. After extracting the region of interest (ROI), we extract the peak temperature, mean temperature, temperature variance and kurtosis as the melt pool temperature field features.

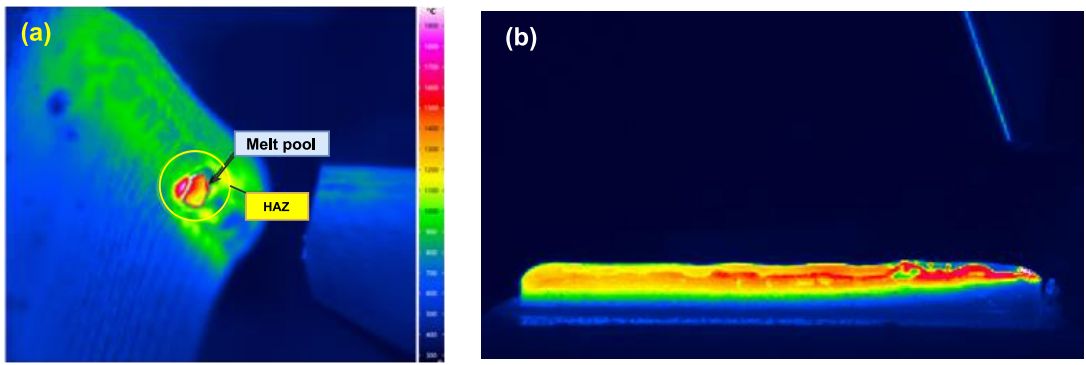

Figure 7. Extraction of molten pool and heat affected zone (HAZ) as the ROI.

### 3.1.4 Surface monitoring

The above-mentioned sensing modalities (acoustic, vision, thermal) provide real-time sensor data feedback, which is useful for online process anomaly detection. However, maintaining dimensional accuracy remains a challenge in the fabrication of large-format metallic parts. Surface defects such as over-built, under-built or geometric distortions need to be identified early in the AM process to avoid further deterioration of the part quality. In-process surface quality inspection with in-situ point cloud processing and adaptive dimension correction was developed in previous work (Chen, Yao, Xu, et al., 2020b). During the process, a laser line scanner performs surface monitoring on a regular basis, as shown in Figure 8. The raw point cloud data with noise and unwanted surfaces is filtered and segmented before being fed into a hybrid ML algorithm to predict the occurrence of surface defects. Based on the extracted geometric deviation boundaries, a robot toolpath is generated with selected laser-on and -off segments to fill in the dent regions, ensuring the dimensional accuracy of the final as-built products.

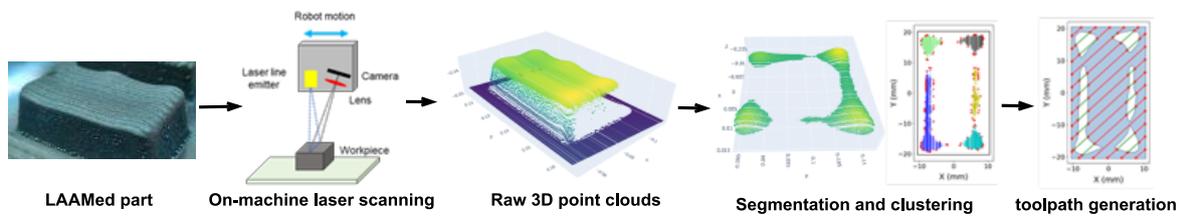

LAAMed part → On-machine laser scanning → Raw 3D point clouds → Segmentation and clustering → toolpath generation

*Figure 8. Rapid surface defect detection and in-process dimension correction.*

### 3.2 Spatiotemporal data fusion for location-dependent quality mapping

After extracting features from the multisensor inputs, they are spatiotemporally fused for location-dependent quality mapping. With the in-house developed ROS-based software platform, all the sensor data features are retrieved and collected simultaneously from the same initial time stamp. Due to the fact that each sensor has a different sampling frequency, multisensor features are resampled to the maximum frequency at 250 Hz (which corresponds to the time axis of robot position data) to avoid information loss. Figure 9 depicts digital twins of the LDED process after the multisensor features are registered spatiotemporally with the robot position data. The interior quality of the part is examined under a microscope, which reveals large keyhole pores in the upper layers, cracks in the middle layers, and a defect-free zone in the lower layers. The multisensor feature values can be mapped with the defect locations. For example, abrupt increases in melt pool width in the middle layer are found to be linked to cracks, while larger melt pools in the top layers generally reflect keyhole porosity conditions. It may also be seen that different sensing modalities have different capabilities in detecting defects. The MFCCs feature from the acoustic signal, for example, misses the evident abnormal signal that corresponds to the cracking area, as can be seen in coaxial melt pool features. Nevertheless, all of the sensor features have followed a similar pattern, with the value increasing (e.g., melt pool geometric features, temperature value) or decreasing (MFCCs) with time, correlating to localised heat build-up. In future study, the spatiotemporally fused multimodal dataset will be used to train machine learning models that predict the defect occurrences at specific locations within the part volume.

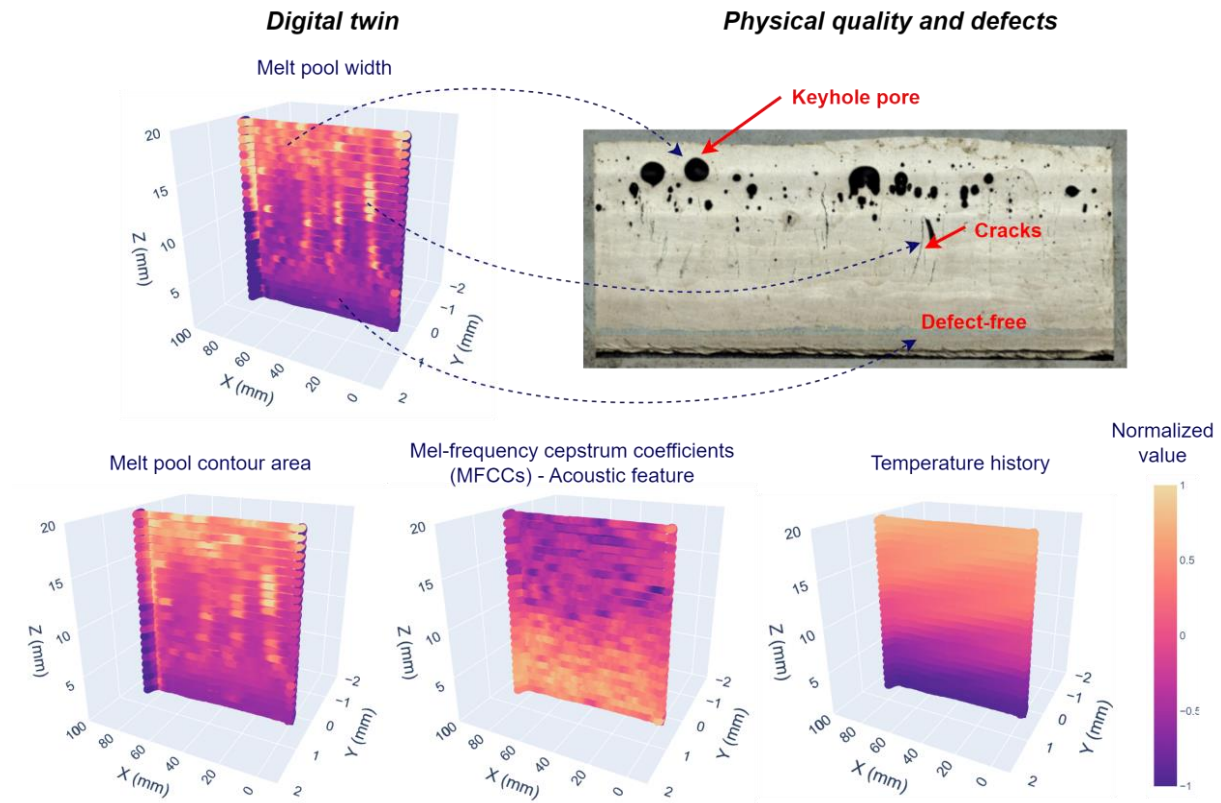

*Figure 9. Multisensor spatiotemporal feature fusion as a digital twin for location-specific quality mapping*

# 4 CONCLUSION AND FUTURE WORK

In this paper, we introduced a multisensor fusion-based digital twin in LDED process. Multisensor features with different sampling frequencies were synchronized and spatiotemporally fused with the robot toolpath positions. The proposed multisensor fusion method provided the foundation for location-specific defect prediction and self-adaptive defect correction. In future work, machine learning models will be developed and trained to predict the defect occurrences such as keyhole porosity and cracking. Robotic machining and LDED toolpath will be automatically generated and executed from the in-house developed software platform, which removes the defects and restores the dimensional accuracy to ensure successful AM production.